\documentclass[aps,prl,superscriptaddress,twocolumn]{revtex4}
\usepackage{graphicx}
\begin{document}
\title{Fabrication method for micro vapor cells for alkali atoms}

\affiliation{5. Physikalisches Institut, Universit\"at Stuttgart, Pfaffenwaldring 57, 70550 Stuttgart}
\affiliation{4. Physikalisches Institut, Universit\"at Stuttgart, Pfaffenwaldring 57, 70550 Stuttgart}
\author{T. Baluktsian}
\affiliation{5. Physikalisches Institut, Universit\"at Stuttgart, Pfaffenwaldring 57, 70550 Stuttgart}
\author{C. Urban}
\affiliation{5. Physikalisches Institut, Universit\"at Stuttgart, Pfaffenwaldring 57, 70550 Stuttgart}
\author{T. Bublat}
\affiliation{4. Physikalisches Institut, Universit\"at Stuttgart, Pfaffenwaldring 57, 70550 Stuttgart}
\author{H. Giessen}
\affiliation{4. Physikalisches Institut, Universit\"at Stuttgart, Pfaffenwaldring 57, 70550 Stuttgart}
\author{R. L\"ow}
\affiliation{5. Physikalisches Institut, Universit\"at Stuttgart, Pfaffenwaldring 57, 70550 Stuttgart}
\author{T. Pfau}
\email[Electronic mail: ]{t.pfau@physik.uni-stuttgart.de}
\affiliation{5. Physikalisches Institut, Universit\"at Stuttgart, Pfaffenwaldring 57, 70550 Stuttgart}

\date{\today}

\begin{abstract}
A quantum network which consists of several components should ideally work on a single physical platform. Neutral alkali atoms have the potential to be very well suited for this purpose due to their electronic structure which involves long lived nuclear spins and very sensitive highly excited Rydberg states. In this paper we describe a fabrication method based on quartz glass to structure arbitrary shapes of microscopic vapor cells. We show that the usual spectroscopic properties known from macroscopic vapor cells are almost unaffected by the strong confinement.
\end{abstract}


\maketitle

Recently a number of quantum devices such as single photon sources, quantum gates and memories, and repeaters have been realized.\cite{obr2009, lvo2009} It is now desirable to accomplish an integrable common technological platform for quantum networks with such components. Based on their electronic structure which involves long lived nuclear spins, very sensitive highly excited Rydberg states and technically convenient optical transitions neutral alkali atoms have the potential to serve for this purpose. Scalable single photon sources can potentially be realized by using the Rydberg blockade.\cite{saf2002} For this a confinement of the atoms on a scale smaller than the blockade radius (typically a few micrometers) is required. Recently it was demonstrated that glass walls do not lead to detrimental dephasing of Rydberg atoms in micrometer-sized vapor cells.\cite{kue2010} Hence they are promising candidates to serve as a scalable single photon source.  Combined with an integrated linear optical network fed by those single photon sources and an integrated quantum memory for those photons based on the long lived nuclear spin in integrable dense thermal vapor cells, all the necessary components for a quantum network can potentially be realized in a holistic approach.\cite{Loe2009}
A crucial technological step is the fabrication and filling of micrometer-sized alkali vapor cells. The miniaturization of alkali vapor cells with a focus on metrology applications has already been brought forward.\cite{kna2005} Another approach is based on the ARROWS technology.\cite{yan2007} Also hollow fibers have been filled with thermal and cold atoms.\cite{gho2006, lig2007, sle2008}\\
In this paper we describe a fabrication method to structure arbitrary shapes of vapor cells with micrometer resolution based on Optical Bonding Of Vacuum Atom Cells (OBOVAC). We show that the filling of these micrometer-sized cells can be carried out and that the usual spectroscopic properties known from larger cells are almost not affected by the confinement. Due to the very convenient optical access and the good optical quality of the microcells the spatial addressing of the cells is straightforward. We use standard absorption and fluorescence spectroscopy methods in a confocal setup to determine the spatial distribution and the optical density of the confined vapor.\\

 \begin{figure}
 \includegraphics[width=8.5cm]{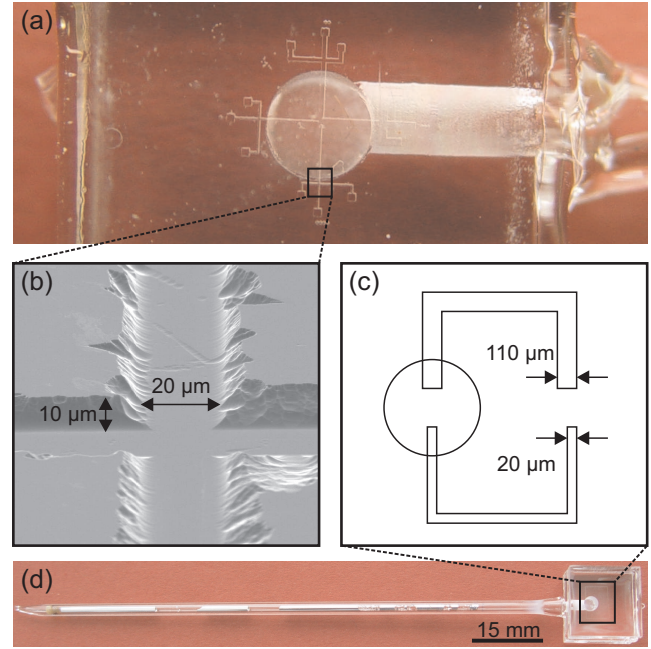}
 \caption{(a) Etched channels and dots ranging from $20$ to $200$ $\mu$m. (b) SEM picture of an etched $20$ $\mu$m trench. (c) Scheme of the structures in the cell shown in (d). (d) Photograph of the spectroscopy cell used for the measurements in this paper. \label{Figure1}}
 \end{figure}

The fabrication process consists of three steps, namely the fabrication of the microstructures, the sealing of the cell, and the filling with rubidium. The vapor cells are made of two quartz substrates with the dimensions $15$ x $15$ x $5$ mm$^3$. The shape of the micrometer-sized structures which provide the confinement for the rubidium vapor is defined by photolithographic methods. First one of the substrates is spin-coated with photoresist which is subsequently exposed using a photomask. The examples shown here are either u-formed paths with widths of $10$ $\mu$m or $100$ $\mu$m or more complex patterns (see Fig. \ref{Figure1}(a)). After the development of the photoresist the exposed areas are removed and the structures are transfered to the substrate by wet etching. The depth of the structures can be varied by changing the etching time. We have chosen the etching time such that we obtain structures with a depth of approximately $10$ $\mu$m. As this depth requires a relatively long etching time the obtained structures were broadened due to the isotropic etching behavior. This resulted in trenches with trapezoidal cross sections ranging from $\sim 20$ to $\sim 40$ $\mu$m and from $\sim 110$ to $\sim 135$ $\mu$m. A scanning electron microscope (SEM) picture of a trench is shown in Figure \ref{Figure1}(b). Closing the trenches results in micrometer-sized channels. Therefore, the structured substrate is covered with a second substrate in optical contact. The covering substrate is prepared with two blind holes so that it is still possible to reach the etched structures from the outside. Subsequently, the two substrates are fused together at the edges of the contact surface. To fill the cell with rubidium a tube is connected to the blind hole. With this tube the whole cell is connected to a vacuum apparatus, which allows us to fill the tube with rubidium under vacuum conditions $\left(< 10^{-6}\text{ mbar}\right)$. Afterwards a liquid droplet of rubidium is filled into the tube and the spectroscopy cell is separated from the vacuum apparatus. The spectroscopy cell thus consists of the two substrates and the tube which serves as rubidium reservoir (Fig. \ref{Figure1}(d)).\\

\begin{figure}
 \includegraphics[width=8.5cm]{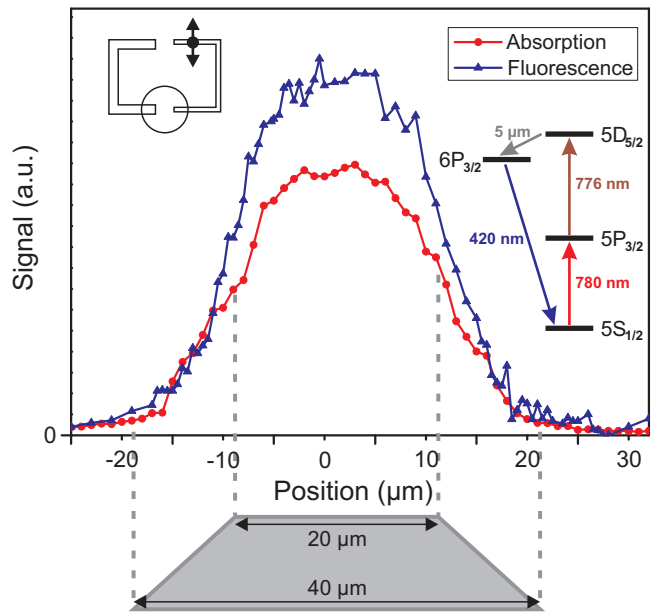}
 \caption{Results of the spatially resolved absorption and fluorescence measurements. The right inset shows a rubidium level scheme with the involved states. In the left inset the position in the cell at which the experiments were performed is indicated. In the lower part a scheme of the cross section of the channel is shown. \label{Figure2}}
 \end{figure}

\begin{figure}
 \includegraphics[width=8.5cm]{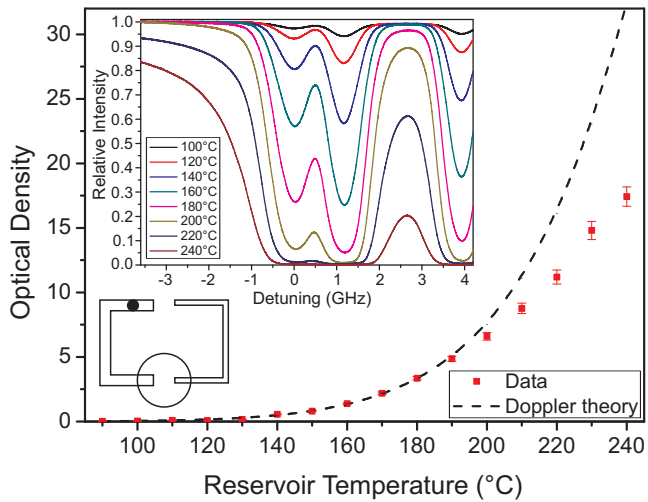}
 \caption{The optical density at the $^{85}$Rb $5S_{1/2}F=3 \rightarrow 5P_{3/2}F=4$ resonance obtained from the experiment is compared to the optical density assuming a Doppler-broadened rubidium vapor (dashed line). The upper inset shows some of the obtained absorption spectra where $0$ detuning is set to the $^{87}$Rb $5S_{1/2}F=2 \rightarrow 5P_{3/2}F=3$ transition. The lower inset indicates the position in the cell at which the measurements were performed. \label{Figure3}}
 \end{figure}

We performed spatially resolved absorption spectroscopy as well as confocal fluorescence spectroscopy to confirm that the fabrication and the filling of the microcell produced by OBOVAC technology was accomplished. We used two stabilized laser diodes, one at $780$ nm and one at $776$ nm. The density of the rubidium atoms in the cell and hence the optical density is adjusted by heating the reservoir. The substrates are always kept at a higher temperature than the reservoir to avoid condensation of rubidium in the microstructures. The lasers pass the cell perpendicular to the contact surface of the substrates, and only $10$ $\mu$m of rubidium vapor are crossed. To be able to resolve structures in the micrometer regime the beams are focused with a microscope objective with a numerical aperture of $0.55$ and a working distance of $13$ mm, which results in a beam width at the focal point of $\sim 3$ $\mu$m. The objective is also used to collect the light emitted by the atoms during the fluorescence measurements. The cell is mounted on a translation stage with an accuracy of $0.5$ $\mu$m and is moved with respect to the fixed position of the focal point to obtain spatially resolved signals. At a cell temperature of $180^\circ$C and a reservoir temperature of $160^\circ$C absorption spectra have been taken at various positions in the spectroscopy cell.
For the fluorescence spectroscopy the laser diode at $780$ nm is locked to the $^{87}$Rb $5S_{1/2}F=2 \rightarrow 5P_{3/2}F=3$ transition, and the laser diode at $776$ nm is locked to the $^{87}$Rb $5P_{3/2}F=3 \rightarrow 5D_{5/2}F=4$ transition for the two step excitation into the $5D_{5/2}$ state as shown in the inset of Figure \ref{Figure2}. Afterwards the atoms decay via the $6P_{3/2}$ state with a fluorescence at $420$ nm back into the ground state, which is analyzed with a spectrometer. The advantage of proving the presence of rubidium with this kind of spectroscopy is its background-free scheme, as the detected light is spectrally well separated from the excitation lasers. At various positions in the spectroscopy cell the fluorescence signal was detected at a cell temperature of $190^\circ$C and a reservoir temperature of $130^\circ$C.
Figure \ref{Figure2} shows the maximum absorption of the $^{87}$Rb $5S_{1/2}F=2 \rightarrow 5P_{3/2}F=1,2,3$ transition as well as the detected fluorescence light at $420$ nm for a scan across the channel with the trapezoidal cross section ranging from $\sim 20$ to $\sim 40$ $\mu$m. As the temperature and thus the rubidium density was kept constant in the experiments both the absorption and the fluorescence signal depend essentially on the vapor layer thickness. The fact that absorption as well as fluorescence itself can be observed shows that the filling of the spectroscopy cell was accomplished. Furthermore the comparison of the spatial dependence of the signals with the trench widths obtained with the SEM confirms that the rubidium is confined to the structures etched into the substrate.\\
Absorption spectra for reservoir temperatures ranging from $90^\circ$C up to $240^\circ$C have been taken to show that the optical density of the thermal rubidium vapor can be varied over a large range. The cell temperature was kept at a $10$ K higher temperature as the reservoir. The $780$ nm laser beam was focused through the $110$ $\mu$m channel with an intensity of $\sim 1/2$ I$_{\text{sat}}$ and probed the $F=2 \rightarrow F^\prime = 1,2,3$ ($^{87}$Rb) and the $F=2,3 \rightarrow F^\prime = 1,2,3,4$ ($^{85}$Rb) transitions of the $D_2$ lines of the two rubidium isotopes. 
The inset in Figure \ref{Figure3} shows some of the spectra. A function assuming a Voigt profile for all the involved transitions was fitted to the data. From the fits the optical density at the $^{85}$Rb $5S_{1/2}F=3 \rightarrow 5P_{3/2}F=4$ resonance was calculated. The results are presented in Figure \ref{Figure3} showing that optical densities of more than $17$ can be reached. For temperatures up to $190^\circ$C the optical density of a thermal vapor only taking into account Doppler broadening and assuming a rubidium density calculated from the vapor-pressure model given by \cite{alc1984} was fitted to the results with the offset temperature as fitting parameter. This leads to a systematic correction of the temperature by $\sim -7$ K which is most likely caused by the temperature sensor being separated a few millimeters from the reservoir. The resulting optical density is also shown in Figure \ref{Figure3}. The discrepancy for higher temperatures results from an increasing linewidth of the Lorentzian contribution to the Voigt profile. This is possibly due to an additional collisional broadening caused by material desorbing from the quartz and will be investigated in further work.
We performed electromagnetically induced transparency (EIT) measurements in the $20$ $\mu$m channel to confirm that the confining walls do not introduce additional line-broadening effects in these cells. For the EIT we used the $5S_{1/2}F=2 \rightarrow 5P_{3/2}F=3$ ($780$ nm) as probe- and the $5P_{3/2}F=3 \rightarrow 5D_{5/2}F=4$ ($776$ nm) as coupling-transition. The probe laser was locked to the $5S_{1/2}F=2 \rightarrow 5P_{3/2}F=3$ transition and its transmission was detected while the coupling laser was scanned across the $5P_{3/2}F=3 \rightarrow 5D_{5/2}F=4$ resonance. We obtained signals with linewidths of $\sim 30$ MHz which is expected as the mean free path is reduced to the size of the channel and due to the averaging of the signal.\\

We introduced the OBOVAC technology which is a method to fabricate arbitrary shapes of vapor cells. Spatially resolved absorption and fluorescence measurements in such a cell demonstrated that the fabrication and the filling of these micrometer-sized cells can be carried out. We also performed EIT measurements which proved that coherent spectroscopy is also possible in these cells. We plan to extend this approach to cell dimensions as small as 1 $\mu$m both in thickness and transverse size. We plan to integrate optical waveguides and test various coatings in future experiments. We envision that the presented technology enables a platform for room temperature quantum networks based on thermal vapor cells.

\begin{acknowledgments}
We acknowledge the assistance and the fruitful discussions with H. K\"ubler, P. Rehme and J. P. Shaffer. We acknowledge the technical assistance of R. August and J. Quack.
\end{acknowledgments}

\end{document}